\documentclass[twocolumn,showpacs,amsmath,amssymb,floatfix,superscriptaddress]{revtex4-1}

\usepackage{graphicx}
\usepackage{dcolumn}
\usepackage{bm}

\usepackage{hyperref}

\begin{document}
	
	\title{Ferromagnetic Damping/Anti-damping in a Periodic 2D Helical surface; A Non-Equilibrium Keldysh Green Function Approach}	
	\author{Farzad Mahfouzi}
	\email{farzad.mahfouzi@gmail.com}
	\affiliation{Department of Physics, California State University, Northridge, California 91330-8268, USA}
	\author{Nicholas Kioussis}
	\affiliation{Department of Physics, California State University, Northridge, California 91330-8268, USA}
	
	\begin{abstract}
		In this paper, we investigate theoretically the spin-orbit torque as well as the Gilbert damping for a two band model of a 2D helical surface state with a Ferromagnetic (FM) exchange coupling. We decompose the density matrix into the Fermi sea and Fermi surface components and obtain their contributions to the electronic transport as well as the spin-orbit torque (SOT). Furthermore, we obtain the expression for the Gilbert damping due to the surface state of a 3D Topological Insulator (TI) and predicted its dependence on the direction of the magnetization precession axis.
	\end{abstract}
	
	\maketitle
	
\section{Introduction}
	The spin-transfer torque (STT) is a phenomenon in which spin current of large enough density injected into a ferromagnetic layer switches its magnetization from one static configuration to another~\cite{Ralph2008}. The origin of STT is absorption of itinerant flow of angular momentum components normal to the magnetization direction. It represents one of the {\em central phenomena} of the second-generation spintronics, focused
	on manipulation of coherent spin states, since reduction of current densities (currently of the order 10$^6$-10$^8$ A/cm$^2$)  required for STT-based magnetization switching is expected to bring commercially viable magnetic random access memory (MRAM)~\cite{Katine2008}. The rich nonequilibrium physics~\cite{Wang2011} arising in the interplay of spin currents carried by fast conduction electrons and collective magnetization dynamics, viewed as the slow classical degree of freedom, is of great fundamental interest.
	
	Very recent experiments~\cite{Miron2010,Liu2012} and theoretical studies~\cite{Manchon2008} have sought STT in nontraditional setups which do not involve the usual two (spin-polarizing and free) F layers with noncollinear magnetizations~\cite{Wang2011}, but rely instead on the spin-orbit coupling (SOC) effects in structures lacking inversion symmetry.  Such ``SO torques''~\cite{Gambardella2011} have been detected~\cite{Miron2010} in Pt/Co/AlO$_x$ lateral devices where current flows in the plane of Co layer. Concurrently, the recent discovery~\cite{Hasan2010} of three-dimensional (3D) topological insulators (TIs), which possess a usual band gap in the bulk while hosting metallic surfaces whose massless Dirac electrons have spins locked with their momenta due to the strong Rashba-type SOC, has led to theoretical proposals to employ these exotic states of matter for spintronics ~\cite{Pesin2012} and STT in particular~\cite{Garate2010}. For example, magnetization of a  ferromagnetic film with perpendicular anisotropy deposited on the TI surface could be switched by interfacial quantum Hall current~\cite{Garate2010}. 

In this paper, we investigate the dynamical properties of a FM/3DTI heterostructure, where the F overlayer covers a TI surface and the device is periodic along in-plane $x-y$ directions. The effect of the F overlayer is a proximity induced exchange field $-{\Delta}_\mathrm{surf} \vec{m} \cdot \vec{\bm \sigma}/2$ superimposed on the Dirac cone dispersion. For a partially covered FM/TI heterostructure, the spin-momentum-locked Dirac electrons flip their spin upon entering into the interface region, thereby inducing a large antidamping-like SOT on the FM~\cite{Qi2008,Mahfouzi2010,Mahfouzi2016}.
The antidamping-like SOT driven by this mechanism which is unique to the surface of TIs has been predicted in Ref.~\cite{Mahfouzi2016}, where a time-dependent nonequilibrium Green function~\cite{Keldysh1965} (NEGF)-based framework was developed. The formalism made it possible to separate different torque components in the presence of arbitrary spin-flip processes within the device. Similar anti-damping torques has also been predicted~\cite{Sinova2014} to exist due to the Berry phase in periodic structures where the device is considered infinite in in-plane directions and a Kubo formula was used to describe the SOT as a linear response to homoginiuos electric field at the interface. However, the connection between the two approaches is not clear and one of the goals of the current paper is to address the similarities and the differences between the two. In the following we present the theoretical formalism of the SOT and damping in the regime of slowly varying parameters of a periodic system in space and time. 
	
Generally, in a quantum system with slowly varying parameters in space and/or time, the system stays close to its equilibrium state ({\it i.e.} adiabatic regime) and the effects of the nonadiabaticity is taken into account perturbatively using adiabatic expansion. Conventionally, this expansion is performed using Wigner representation~\cite{Haug2008} after the separation of the fast and slow variations in space and/or time.~\cite{Bode2012} The slow variation implies that the NEGFs vary slowly with the central space ( time ), \mbox{$\vec{x}_c=(\vec{x}+\vec{x}')/2$} \mbox{( $t_c=(t+t')/2$ )}, while they change fast with
the relative space (time), \mbox{$\vec{x}_r=\vec{x}-\vec{x}'$} \mbox{( $t_r=t-t'$ )}. Here we use an alternative approach, where we consider $(x,t)$ and \mbox{$(\vec{x}_r,t_r)$} as the natural variables to describe the close to adiabatic apace-time evolution of NEGFs and then perform the following Fourier transform
\begin{equation}\label{eq:ft}
\check{\mathbf{G}}(\vec{x}t;\vec{x}'t')=\int \frac{dE}{2\pi}\frac{d\vec{k}}{\Omega_k} e^{iE(t-t')+i\vec{k}\cdot(\vec{x}-\vec{x}')} \check{\mathbf{G}}_{\vec{k}E}(\vec{x}t).
\end{equation}
where, $\Omega_k$ is the volume of the phase space that the $\vec{k}$-integration is being performed. The standard Dyson equation of motion for $\check{\mathbf{G}}(\vec{x}t;\vec{x}'t')$ is cumbersome to manipulate~\cite{Arrachea2007,Jauho1994} or solve numerically,~\cite{Gaury2014} so they are usually transformed to some other representation.\cite{Mahfouzi2012}  
	Generalizing the equation to take into account slowly varying time and spatial dependence of the Hamiltonian we obtain,
	\begin{align}\label{eq:KeldyshEqAdia}
	&\check{\bold{G}}=
	\left( \begin{array}{cc}
	\bold{G}^{r} & \bold{G}^{<} \\
	0 & \bold{G}^{a} \end{array} \right),\\
	&=\left( \begin{array}{cc}
	\bold{G}_{ad}^{r,-1}-i\boldsymbol{\mathcal{D}}_{xt} & \bold{\Sigma}^< \\
	0 & \bold{G}_{ad}^{a,-1}-i\boldsymbol{\mathcal{D}}_{xt} \end{array} \right)^{-1},\nonumber
	\end{align}
	where,
	\begin{subequations}
	\begin{align}\label{eq:GrAdia}
	&\bold{G}_{ad}^{r,-1}=(E-i\eta)\bold{1}-\bold{H}(\vec{k},t)-\mu(\vec{x}),\\
	&\bold{\Sigma}^<=-2i\eta f(E-i\frac{\partial}{\partial t}-\mu(\vec{x})),\\
	&\boldsymbol{\mathcal{D}}_{xt}=\frac{\partial}{\partial t}+\frac{\partial \bold{H}}{\partial \vec{k}}\cdot\vec{\nabla},
	\end{align}
	\end{subequations}
	and, $\eta=\hbar/2\tau$ is the phenomenological broadening parameter, where $\tau$ is the relaxation time. It is worth mentioning that for a finite $\eta$ the number of particles is not conserved, and a more accurate interpretation of the introduced broadening might be to consider it as an energy-independent scape rate of electrons to fictitious reservoirs attached to the positions $\vec{x}$. Consequently, a finite broadening could be interpreted as the existence of an interface in the model between each atom in the system and the reservoir that is spread homogeneously along the infinite periodic system. 
	
	Eq. \eqref{eq:KeldyshEqAdia} shows that the effect of the space/time variation is to replace $E\rightarrow E-i\partial/\partial t$ and $\vec{k}\rightarrow \vec{k}-i\vec{\nabla}$ in the equation of motion for the GFs in stationary state. To the lowest order with respect to the derivatives we can write,
	\begin{align}\label{eq:KeldyshEqAdiaSol}
	\check{\bold{G}}=\check{\bold{G}}_{ad}-i\frac{\partial\check{\bold{G}}_{ad}}{\partial E}\frac{\partial \check{\bold{G}}^{-1}_{ad}}{\partial t}\check{\bold{G}}_{ad}-i\frac{\partial\check{\bold{G}}_{ad}}{\partial \vec{k}}\cdot\vec{\nabla}\check{\bold{G}}^{-1}_{ad}\check{\bold{G}}_{ad},
	\end{align}
	where,
	\begin{align}
	\check{\bold{G}}_{ad}^{-1}=
	\left( \begin{array}{cc}
	\bold{G}_{ad}^{r,-1} &  -2i\eta f(E-\mu(\vec{x})) \\
	0 & \bold{G}_{ad}^{a,-1} \end{array} \right).
	\end{align}
	For the density matrix of the system, $\boldsymbol{\rho}(t)=\frac{1}{i}\bold{G}^<(t,t)$, we obtain,
	
	\begin{align}
	\boldsymbol{\rho}_{\vec{k},t}^{neq}\approx -\int \frac{dE}{2\pi}\Re\left(\left[\mathcal{D}(\bold{G}_{ad}^{r}),\bold{G}_{ad}^{r}\right]f
	+2i\eta\mathcal{D}(\bold{G}_{ad}^{r})\bold{G}_{ad}^{a}\frac{\partial f}{\partial E}\right)\label{eq:DM_xt}
	\end{align}
	where $\mathcal{D}=\frac{\partial}{\partial t}-\vec{\nabla}\mu\cdot\frac{\partial}{\partial \vec{k}}$ is the differential operator acting on the slowly varying parameters in space and time. The details of the derivation is presented in Appendix.\ref{App_a}. The density matrix in Eq. \eqref{eq:DM_xt} is the central formula of the paper and consists of two terms; the first term contains the equilibrium Fermi distribution function from the electrons bellow the Fermi surface occupying a slowly (linearly) varying single particle states that has only interband contributions and can as well be formulated in terms of the Berry phase as we will show the following sections, and; the second term corresponds to the electrons with Fermi energy (at zero temperature we have, $\partial f/\partial E=\delta(E-E_F)$) which are the only electrons allowed to get excited in the presence of the slowly varying perturbations. The fact that the first term originates from the assumption that the electric field is constant inside the metallic FM suggests that this term might disappear once the screening effect is included. On the other hand, due to the fact that the second term corresponds to the nonequilibrium electrons injected from the fictitious reservoirs attached to the device through the scape rate $\eta$, it might capture the possible physical processes that occur at the contact region and makes it more suitable for the calculation of the relevant physical observables in such systems.
	
	Using the expression for the nonequilibrium density matrix the local spin density can be obtained from,
	\begin{align}\label{eq:TrqExpr2}
	\vec{S}_{neq}(t)=\langle\vec{\sigma}\rangle_{neq}=\frac{1}{4\pi^2} \int d^2\vec{k}Tr[\bold{\rho}_{\vec{k},t}^{neq}\vec{\sigma}],
	\end{align}
	where $\langle...\rangle_{neq}$ refers to the ensemble average over many-body states out of equilibrium demonstrated by the nonequilibrium density matrix of the electrons and, $Tr$ refers to the trace.
	In this case the time derivative in the differential operator $\mathcal{D}$ leads to the damping of the dynamics of the ferromagnet while the momentum derivative leads to either damping or anti-damping of the FM dynamics depending on the direction of the applied electric field. In the following section we apply the formalism to a two band helical surface state model attached to a FM. 

\section{SOT and Damping of a Helical 2D Surface}

A two band Hamiltonian model for the system can be generally written as,
\begin{align}\label{eq:2D_Hamil}
\bold{H}(\vec{k},t)=\varepsilon_0(\vec{k})\boldsymbol{1}+\vec{h}(\vec{k},t)\cdot\vec{\boldsymbol{\sigma}}
\end{align}
where, $\vec{h}=\vec{h}_{so}(\vec{k})+\frac{\Delta_{xc}(\vec{k})}{2}\vec{m}(t)$, with $\vec{h}_{so}(\vec{k})=-\vec{h}_{so}(-\vec{k})$ and $\Delta_{xc}(\vec{k})=\Delta_{xc}(-\vec{k})$ being spin-orbit and magnetic exchange coupling terms respectively. In particular in the case of Rashba type helical states we have $\vec{h}_{so}=\alpha_{so}\hat{e}_z\times\vec{k}$.
In this case for the adiabatic single particle GF we have,
\begin{align}\label{eq:ad_GF}
	\bold{G}_{ad}^{r}(E,t)=\frac{(E-\varepsilon_0-i\eta)\boldsymbol{1}+\vec{h}\cdot\vec{\boldsymbol{\sigma}}}{(E-\varepsilon_0-i\eta)^2-|\vec{h}|^2}
\end{align}
From Eq.~\eqref{eq:TrqExpr2} for the local spin density, we obtain (See Appendix \ref{App_b} for details),
\begin{align}\label{eq:neqLocSp1}
\vec{S}_{neq}(t)=&\int \frac{d^2\vec{k}}{4\pi^2}\left(\frac{\vec{h}\times\mathcal{D}\vec{h}}{2|\vec{h}|^3}\left(f_1-f_2\right)-\frac{(\vec{\nabla}\mu\cdot\vec{v}_0)\vec{h}}{2\eta|\vec{h}|}\left(f'_1-f'_2\right)\right.\nonumber\\
&\left.+(\frac{\vec{h}\times\mathcal{D}\vec{h}}{2|\vec{h}|^2}+\frac{\eta\mathcal{D}\vec{h}-\frac{1}{\eta}(\vec{h}\cdot\mathcal{D}\vec{h})\vec{h}}{2|\vec{h}|^2})\left(f'_1+f'_2\right)\right)
\end{align}
where, $f_{1,2}=f(\varepsilon_0\pm|\vec{h}|)$ and $\vec{v}_0=\partial\varepsilon_0/\partial\vec{k}$ is the group velocity of electrons in the absence of the SOI. Here, we assume $\eta\ll|\vec{h}|$ which corresponds to a system close to the ballistic regime. In this expression we kept the $\eta\mathcal{D}\vec{h}$ because of its unique vector orientation characteristics. As it becomes clear in the following, the first term in Eq.~\eqref{eq:neqLocSp1} is a topological quantity which in the presence of an electric field becomes dissipative and leads to an anti-damping torque. The second term in this expression leads to the Rashba-Edelstein field-like torque which is a nondissipative observable. The third term has the exact form as the first term with the difference that it is strictly a Fermi surface quantity. The fourth term, also leads to a field like torque that as we will see in the following has similar features as the Rashba-Edelstein effect. It is important to pay attention that unlike the first term, the rest of the terms in Eq.~\eqref{eq:neqLocSp1} are solely due to the flow of the non-equilibrium electrons on the Fermi surface. Furthermore, we notice that the terms that lead to dissipation in the presence of an electric field ($\mathcal{D}\equiv\vec{\nabla}\mu\cdot\frac{\partial}{\partial \vec{k}}$) become nondissipative when we consider $\mathcal{D}\equiv\partial/\partial t$  and vice versa.
\subsection{Surface State of a 3D-TI} 
In the case of the surface state of  a 3D-TI, as an approximation we can ignore $\varepsilon_0(\vec{k})$ and consider the helical term as the only kinetic term of the Hamiltonian. In this case the local charge current and the nonequilibrium local spin density share a similar expression, $\vec{I}=\langle\partial (\vec{h}\cdot\vec{\sigma})/\partial \vec{k}\rangle$. For the conductivity, analogous to  Eq.~\eqref{eq:neqLocSp1}, we obtain, 
\begin{align}\label{eq:LocCur}
\sigma_{ij}=e\int &\frac{d^2\vec{k}}{4\pi^2}\left(\frac{\vec{h}\cdot\frac{\partial\vec{h}}{\partial k_i}\times\frac{\partial\vec{h}}{\partial k_j}}{2|\vec{h}|^2}\left(\frac{f_1-f_2}{|\vec{h}|}+f'_1+f'_2\right)\delta_{i\neq j}\right.\nonumber\\
&\left.+\frac{-\eta|\frac{\partial\vec{h}}{\partial k_i}|^2+\frac{1}{\eta}(\frac{\partial|\vec{h}|^2}{\partial k_i})^2}{2|\vec{h}|^2}\left(f'_1+f'_2\right)\delta_{ij}\right)
\end{align} 
This shows that the Fermi sea component of the density matrix contributes only to the anomalous Hall conductivity which is in terms of a winding number. On the other hand, the second term is finite only for the longitudinal components of the conductivity and can be rewritten in terms of the group velocity of the electrons in the system which leads to the Drude-like formula.

Should the linear dispersion approximation for the kinetic term in the Hamiltonian be valid in the range of the energy scale corresponding to the magnetic exchange coupling $\Delta_{xc}$ (i.e. when $v_F\gg\Delta_{xc}$), the effect of the in-plane component of the magnetic exchange coupling is to shift the Dirac point (i.e. center of the k-space integration) which does not affect the result of the k-integration. In this case after performing the partial time-momentum derivatives, ($\mathcal{D}(\vec{h})=\frac{\Delta_{xc}}{2}\frac{\partial \vec{m}}{\partial t}-v_F\hat{e}_z\times\vec{\nabla}\mu$), we use $\vec{h}(\vec{k},t)=v_{F}\hat{e}_z\times\vec{k}+\frac{\Delta_{xc}}{2}m_z(t)\hat{e}_z$, to obtain,
\begin{align}\label{eq:neqLocSp2}
\vec{S}_{neq}&(t)=\int \frac{kd{k}}{4\pi|\vec{h}|^2}\left(\vec{S}_{1}\frac{f_1-f_2}{|\vec{h}|}+(\vec{S}_{1}+\vec{S}_{2})\left(f'_1+f'_2\right)\right),
\end{align}
where,
\begin{align}
\vec{S}_1(\vec{k},t)=&\frac{\Delta^2_{xc}}{4}m_z(t)\hat{e}_z\times\frac{\partial \vec{m}}{\partial t}+\frac{\Delta_{xc}v_F}{2}m_z(t)\vec{\nabla}\mu\label{eq:neqLocSp3a}\\
\vec{S}_2(\vec{k},t)=&\frac{\Delta_{xc}}{4\eta}(2\eta^2-v_F^2|k|^2)(\frac{\partial m_x}{\partial t}\hat{e}_x+\frac{\partial m_y}{\partial t}\hat{e}_y)\nonumber\\
&+\frac{\Delta_{xc}}{4\eta}(2\eta^2-\frac{\Delta_{xc}^2m_z^2}{2})\frac{\partial {m}_z}{\partial t}\hat{e}_z\nonumber\\
&-\frac{v_F}{\eta}(\eta^2 -\frac{v_F^2|k|^2}{2})\hat{e}_z\times\vec{\nabla}\mu\label{eq:neqLocSp3b}
\end{align}
The dynamics of the FM obeys the LLG equation where the conductions electrons insert torque on the FM moments through the magnetic exchange coupling,
\begin{align}\label{eq:LLG1}
\frac{\partial \vec{m}}{\partial t}=\vec{m}\times\left(\gamma\vec{B}_{ext}+\frac{\Delta_{xc}}{2}\vec{S}_{neq}(t)-\sum_{ij}\alpha^{ij}_0\frac{\partial m_i}{\partial t}\hat{e}_{j}\right)
\end{align}
where, $\alpha^{ij}_0=\alpha^{ji}_0$, with $i,j=x,y,z$, is the intrinsic Gilbert damping tensor of the FM in the absence of the TI surface state and $\vec{B}_{ext}$ is the total magnetic field applied on the FM aside from the contribution of the nonequilibrium electrons.

While the terms that consist of $\vec{\nabla}\mu$ are called SOT, the ones that contain $\frac{\partial \vec{m}}{\partial t}$ are generally responsible for the damping of the FM dynamics. However, we notice that $\hat{e}_z\times\frac{\partial \vec{m}}{\partial t}$ term in Eq.~\eqref{eq:neqLocSp3a} which arises from the Berry curvature, becomes $m_z\frac{\partial \vec{m}}{\partial t}$ in the LLG equation that does not contribute to the damping and only renormalizes the coefficient of the left hand side of the Eq.~\eqref{eq:LLG1}. The second term in the Eq.~\eqref{eq:neqLocSp3a}, is the anti-damping SOT pointing along ($e_z\times\vec{\nabla}\mu$)-axis. The cone angle dependence of the anti-damping term can be checked by assuming an electric field along the $x$-axis when the FM precesses around the $y$-axis, (i.e. $\vec{m}(t)=\cos(\theta)\hat{e}_y+\sin(\theta)\cos(\omega t)\hat{e}_x+\sin(\theta)\sin(\omega t)\hat{e}_z$). In this case the average of the SOT along the $y$-axis in one period of the precession leads to the average of the antidamping SOT that shows a $\sin^2(\theta)$ dependence, which is typical for the damping-like torques. Keeping in mind that in this section we consider $v_F\gg \Delta_{xc}$, the first and second terms in Eq.~\eqref{eq:neqLocSp3b} show that the Gilbert damping increases as the precession axis goes from in-plane ($x$ or $y$) to out of plane ($z$) direction. Furthermore, when the precession axis is in-plane (e.g. along $y$-axis), the damping rate due to the oscillation of the out of plane component of the magnetization ($\partial m_z/\partial t$) has a $\sin^4(\theta)$ dependence that can be ignored for low power measurement of the Gilbert damping $\theta\ll 1$. This leaves us with the contribution from the in-plane magnetization oscillation  ($\partial m_x\partial t$) only. Therefore, the Gilbert damping for in-plane magnetization becomes half of the case when magnetization is out-of-plane. The anisotropic dependence of the Gilbert damping can be used to verify the existence of the surface state of the 3DTI as well as the proximity induce magnetization at the interface between a FM and a 3DTI. Finally, the third term in Eq.~\eqref{eq:neqLocSp3b} demonstrates a field like SOT with the same vector field characteristics as the Rashba-Edelstein effect.

\section{Conclusion}

	In conclusion, we have developed a linear response NEGF framework which provides unified treatment of 
	both spin torque and damping due to SOC at interfaces. We obtained the expressions for both damping and anti-damping torques in the presence of a linear gradiance of the electric field and adiabatic time dependence of the magnetization dynamics for a helical state corresponding to the surface state of a 3D topological insulator. We present the exact expressions for the damping/anti-damping SOT as well as the field like torques and showed that, ({\it i}); Both Fermi surface and Fermi sea contribute similarly to the anti-damping SOT as well as the Hall conductivity and, ({\it ii}); The Gilbert damping due to the surface state of a 3D TI when the magnetization is in-plane is less than the Gilbert damping when it is in the out-of-plane direction. This dependence can be used as a unique signature of the helicity of the surface states of the 3DTIs and the presence of the proximity induced magnetic exchange from the FM overlayer.

\begin{acknowledgments}
We thank Branislav K. Nikoli\'c for the fruitful discussions. F. M. and N. K. were supported by NSF PREM Grant No. 1205734.
\end{acknowledgments}
		\appendix
		\section{Derivation of the Density Matrix}\label{App_a}
		Using Eqs.~\eqref{eq:KeldyshEqAdia} and .~\eqref{eq:KeldyshEqAdiaSol} it is straightforward to obtain,
		\begin{align}
		\bold{G}^{<}=& (\bold{G}_{ad}^{r}-\bold{G}_{ad}^{a})f-2\eta f'\nabla\mu\cdot\frac{\partial \bold{G}_{ad}^{r}}{\partial \boldsymbol{k}}\bold{G}_{ad}^{a}\nonumber\\
		&+i\frac{\partial \bold{G}_{ad}^{<}}{\partial E}\frac{\partial\bold{H}}{\partial t}\bold{G}_{ad}^{a}+i\frac{\partial \bold{G}_{ad}^{r}}{\partial E}\frac{\partial\bold{H}}{\partial t}\bold{G}_{ad}^{<}\nonumber\\
		&+i\frac{\partial \bold{G}_{ad}^{<}}{\partial \boldsymbol{k}}\cdot\nabla\bold{H}\bold{G}_{ad}^{a}+i\frac{\partial \bold{G}_{ad}^{r}}{\partial \boldsymbol{k}}\cdot\nabla\bold{H}\bold{G}_{ad}^{<}
		\end{align}
		We plug in the expression for the adiabatic lesser GF in equilibrium, $\bold{G}_{ad}^<=2i\eta f\bold{G}_{ad}^r\bold{G}_{ad}^a=(\bold{G}_{ad}^r-\bold{G}_{ad}^a)f$, and obtain,
		\begin{align}
		\bold{G}^{<}&= (\bold{G}_{ad}^{r}-\bold{G}_{ad}^{a})f-2\eta f'\nabla\mu\cdot\frac{\partial \bold{G}_{ad}^{r}}{\partial \boldsymbol{k}}\bold{G}_{ad}^{a}\nonumber\\
		&+if\frac{\partial (\bold{G}_{ad}^{r}-\bold{G}_{ad}^{a})}{\partial E}\frac{\partial\bold{H}}{\partial t}\bold{G}_{ad}^{a}
		+if\frac{\partial \bold{G}_{ad}^{r}}{\partial E}\frac{\partial\bold{H}}{\partial t}(\bold{G}_{ad}^{r}-\bold{G}_{ad}^{a})\nonumber\\
		&+if'( \bold{G}_{ad}^{r}-\bold{G}_{ad}^{a})\frac{\partial\bold{H}}{\partial t}\bold{G}_{ad}^{a}
		+if\frac{\partial (\bold{G}_{ad}^{r}-\bold{G}_{ad}^{a})}{\partial \boldsymbol{k}}\cdot\nabla\bold{H}\bold{G}_{ad}^{a}\nonumber\\
		&+if\frac{\partial \bold{G}_{ad}^{r}}{\partial \boldsymbol{k}}\cdot\nabla\bold{H}(\bold{G}_{ad}^{r}-\bold{G}_{ad}^{a}).
		\end{align}
		Expanding the terms, leads to,
		\begin{align}
		\bold{G}^{<}=& \left(\bold{G}_{ad}^{r}-\bold{G}_{ad}^{a}+i\bold{G}_{ad}^{a}\frac{\partial\boldsymbol{H}(t)}{\partial t}\frac{\partial \bold{G}_{ad}^{a}}{\partial E}\right.\nonumber\\
		&-i\bold{G}^{r}_{ad}\frac{\partial \boldsymbol{H}}{\partial \boldsymbol{k}}\cdot\nabla\mu(\boldsymbol{x})\frac{\partial \bold{G}^{r}_{ad}}{\partial E}-i\frac{\partial \bold{G}_{ad}^{a}}{\partial E}\frac{\partial\boldsymbol{H}(t)}{\partial t}\bold{G}_{ad}^{a}\nonumber\\
		&\left.+i\frac{\partial \bold{G}^{r}_{ad}}{\partial E}\frac{\partial \boldsymbol{H}}{\partial \boldsymbol{k}}\cdot\nabla\mu(\boldsymbol{x})\bold{G}^{r}_{ad}\right)f\nonumber\\
		&+i f'\bold{G}_{ad}^{r}\nabla\mu\cdot\frac{\partial \boldsymbol{H}}{\partial \boldsymbol{k}}(\bold{G}_{ad}^{r}-\bold{G}_{ad}^{a})\nonumber\\
		&+if'( \bold{G}_{ad}^{r}-\bold{G}_{ad}^{a})\frac{\partial\bold{H}}{\partial t}\bold{G}_{ad}^{a},
		\end{align}
		where, for the first and third lines we have used their anti-Hermitian forms instead. Since to calculate the density matrix we integrate $\bold{G}^<$ over energy, we can use integration by parts and obtain,
		\begin{align}
		\bold{G}^{<}&= \left(\bold{G}_{ad}^{r}-\bold{G}_{ad}^{a}+2i\bold{G}_{ad}^{a}\frac{\partial\boldsymbol{H}(t)}{\partial t}\frac{\partial \bold{G}_{ad}^{a}}{\partial E}\right.\nonumber\\
		&\left.-2i\bold{G}^{r}_{ad}\frac{\partial \boldsymbol{H}}{\partial \boldsymbol{k}}\cdot\nabla\mu(\boldsymbol{x})\frac{\partial \bold{G}^{r}_{ad}}{\partial E}\right)f\nonumber\\
		&-i f'\left(\bold{G}_{ad}^{r}\nabla\mu\cdot\frac{\partial \boldsymbol{H}}{\partial \boldsymbol{k}}\bold{G}_{ad}^{a}- \bold{G}_{ad}^{r}\frac{\partial\bold{H}}{\partial t}\bold{G}_{ad}^{a}\right)\nonumber\\
		&=(\bold{G}_{ad}^{r}-\bold{G}_{ad}^{a})f\nonumber\\
		&+i \left(2\bold{G}_{ad}^{r}\mathcal{D}(\bold{H})\frac{\partial\bold{G}_{ad}^{r}}{\partial E}f+
		\bold{G}_{ad}^{r}\mathcal{D}(\bold{H})\bold{G}_{ad}^{a}f'\right).
		\end{align}	
		where we define, $\mathcal{D}=\frac{\partial}{\partial t}-\nabla\mu\cdot\frac{\partial }{\partial \boldsymbol{k}}$. 
		Finally, using the identity,
		\begin{align}
		2\bold{G}_{ad}^{r}\mathcal{D}(\bold{H})\frac{\partial\bold{G}_{ad}^{r}}{\partial E}=&\bold{G}_{ad}^{r}\mathcal{D}(\bold{H})\frac{\partial\bold{G}_{ad}^{r}}{\partial E}-\frac{\partial\bold{G}_{ad}^{r}}{\partial E}\mathcal{D}(\bold{H})\bold{G}_{ad}^{r}\nonumber\\
		&+\frac{\partial\left(\bold{G}_{ad}^{r}\mathcal{D}(\bold{H})\bold{G}_{ad}^{r}\right)}{\partial E},
		\end{align}	
		and performing the differential over the energy, we arrive at Eq.~\eqref{eq:DM_xt}.
	\section{Derivation of the Local Spin Density}\label{App_b}
	\begin{widetext}
		From Eqs.~\eqref{eq:DM_xt} and \eqref{eq:TrqExpr2}, the local spin density can be written as, $\vec{S}_{neq}=\vec{S}^{(1)}_{neq}+\vec{S}^{(2)}_{neq}$, where $\vec{S}^{(1)}_{neq}$ is due to the nonequilibrium electrons at the Fermi surface and for a two band model can be calculated as the following,
		\begin{align}\label{eq:LocSp_1}
		\vec{S}^{(1)}_{neq}\approx & -2\eta\Im\int \frac{dE}{2\pi}Tr\left[\frac{\left(\mathcal{D}(\vec{h}\cdot\vec{\boldsymbol{\sigma}})+\vec{\nabla}\mu\cdot\vec{v}_0\boldsymbol{1}\right)\left((E-\varepsilon_0+i\eta)\boldsymbol{1}+\vec{h}\cdot\vec{\boldsymbol{\sigma}}\right)}{((E-\varepsilon_0-i\eta)^2-|\vec{h}|^2)((E-\varepsilon_0+i\eta)^2-|\vec{h}|^2)}\vec{\boldsymbol{\sigma}}\right.\nonumber\\
		& \left.+\mathcal{D}(\frac{1}{(E-\varepsilon_0-i\eta)^2-|\vec{h}|^2})\frac{((E-\varepsilon_0+i\eta)\boldsymbol{1}+\vec{h}\cdot\vec{\boldsymbol{\sigma}})\vec{\boldsymbol{\sigma}}((E-\varepsilon_0-i\eta)\boldsymbol{1}+\vec{h}\cdot\vec{\boldsymbol{\sigma}})}{(E-\varepsilon_0+i\eta)^2-|\vec{h}|^2}\right]f'(E)
		\end{align}
		Preforming the trace over the Pauli matrix, we obtain,
		\begin{align}
		\vec{S}^{(1)}_{neq}\approx  -4\eta\int \frac{dE}{2\pi}&\left(\frac{\eta\mathcal{D}\vec{h}+\mathcal{D}(\vec{h})\times\vec{h}}{((E-\varepsilon_0-i\eta)^2-|\vec{h}|^2)(E-\varepsilon_0+i\eta)^2-|\vec{h}|^2)}\right.\nonumber\\
		&\left.+4\Im(\frac{\left(\vec{h}\cdot\mathcal{D}(\vec{h})+(E-\varepsilon_0-i\eta)(\vec{\nabla}\mu\cdot\vec{v}_0)\right)(E-\varepsilon_0)\vec{h}}{((E-\varepsilon_0-i\eta)^2-|\vec{h}|^2)^2((E-\varepsilon_0+i\eta)^2-|\vec{h}|^2)})\right)f'(E)
		\end{align}
		In the limit of small broadening, $\eta\ll |\vec{h}|$, we obtain,
		\begin{align}
		\vec{S}^{(1)}_{neq}\approx   &-\frac{\eta\mathcal{D}\vec{h}-\vec{h}\times\mathcal{D}\vec{h}+\frac{1}{\eta}\vec{h}\cdot\mathcal{D}(\vec{h})\vec{h}}{2|\vec{h}|^2}\left(f'(\varepsilon_0+|\vec{h}|)+f'(\varepsilon_0-|\vec{h}|)\right)\nonumber\\
		&-\frac{1}{\eta}(\vec{\nabla}\mu\cdot\vec{v}_0)\frac{\vec{h}}{2|\vec{h}|}\left(f'(\varepsilon_0+|\vec{h}|)-f'(\varepsilon_0-|\vec{h}|)\right)
		\end{align}
		For the Fermi sea contribution to the nonequilibrium local spin density we have,
		\begin{align}\label{eq:LocSp_2}
		\vec{S}^{(2)}_{neq}\approx & \Re\int \frac{dE}{2\pi}Tr\left[\frac{\partial}{\partial E}\left(\frac{(E-\varepsilon_0-i\eta)\boldsymbol{1}+\vec{h}\cdot\vec{\boldsymbol{\sigma}}}{(E-\varepsilon_0-i\eta)^2-|\vec{h}|^2}\right)\left(\mathcal{D}(\vec{h}\cdot\vec{\boldsymbol{\sigma}})+\vec{\nabla}\mu\cdot\vec{v}_0\boldsymbol{1}\right)\frac{(E-\varepsilon_0-i\eta)\boldsymbol{1}+\vec{h}\cdot\vec{\boldsymbol{\sigma}}}{(E-\varepsilon_0-i\eta)^2-|\vec{h}|^2}\vec{\boldsymbol{\sigma}}\right.\nonumber\\
		&\left.-\frac{(E-\varepsilon_0-i\eta)\boldsymbol{1}+\vec{h}\cdot\vec{\boldsymbol{\sigma}}}{(E-\varepsilon_0-i\eta)^2-|\vec{h}|^2}\left(\mathcal{D}(\vec{h}\cdot\vec{\boldsymbol{\sigma}})+\vec{\nabla}\mu\cdot\vec{v}_0\boldsymbol{1}\right)\frac{\partial}{\partial E}\left(\frac{(E-\varepsilon_0-i\eta)\boldsymbol{1}+\vec{h}\cdot\vec{\boldsymbol{\sigma}}}{(E-\varepsilon_0-i\eta)^2-|\vec{h}|^2}\right)\vec{\boldsymbol{\sigma}}\right]f(E).
		\end{align}
		Similarly, we obtain,
		\begin{align}
		\vec{S}^{(2)}_{neq}\approx & \Re\int \frac{dE}{2\pi}Tr\left[\frac{\left[\mathcal{D}(\vec{h}\cdot\vec{\boldsymbol{\sigma}})+\vec{\nabla}\mu\cdot\vec{v}_0\boldsymbol{1},(E-\varepsilon_0-i\eta)\boldsymbol{1}+\vec{h}\cdot\vec{\boldsymbol{\sigma}}\right]}{((E-\varepsilon_0-i\eta)^2-|\vec{h}|^2)^2}\vec{\boldsymbol{\sigma}}\right]f(E)\nonumber\\
		\approx & 2\Im\int \frac{dE}{\pi}\frac{\mathcal{D}(\vec{h})\times\vec{h}}{((E-\varepsilon_0-i\eta)^2-|\vec{h}|^2)^2}f(E)\nonumber\\
		\approx & -\frac{1}{2}\frac{\mathcal{D}(\vec{h})\times\vec{h}}{|\vec{h}|^3}\left(f(\varepsilon_0+|\vec{h}|)-f(\varepsilon_0-|\vec{h}|)\right)
		\end{align}
	\end{widetext}
		
	

\begin{thebibliography}{10}
	
	\bibitem{Ralph2008}
	D. Ralph and M. Stiles, Journal of Magnetism and Magnetic Materials {\bf 320},
	  1190  (2008).
	
	\bibitem{Katine2008}
	J. Katine and E.~E. Fullerton, Journal of Magnetism and Magnetic Materials {\bf
	  320},  1217  (2008).
	
	\bibitem{Wang2011}
	C. Wang {\it et~al.}, Nature Physics {\bf 7},  496  (2011).
	
	\bibitem{Miron2010}
	I.~M. Miron {\it et~al.}, Nature Materials {\bf 9},  230  (2010).
	
	\bibitem{Liu2012}
	L. Liu {\it et~al.}, Phys. Rev. Lett. {\bf 109},  096602  (2012).
	
	\bibitem{Manchon2008}
	A. Manchon and S. Zhang, Physical Review B {\bf 78},  212405  (2008).
	
	\bibitem{Gambardella2011}
	P. Gambardella and I.~M. Miron, Philos. Transact. A Math. Phys. Eng. Sci. {\bf
	  369},  3175  (2011).
	
	\bibitem{Hasan2010}
	M.~Z. Hasan and C.~L. Kane, Rev. Mod. Phys. {\bf 82},  3045  (2010).
	
	\bibitem{Pesin2012}
	D. Pesin and A.~H. MacDonald, Nature Mater. {\bf 11},  409  (2012).
	
	\bibitem{Garate2010}
	I. Garate and M. Franz, Phys. Rev. Lett. {\bf 104},  146802  (2010).
	
	\bibitem{Mahfouzi2012}
	F. Mahfouzi, J. Fabian, N. Nagaosa, and B.~K. Nikoli\ifmmode~\acute{c}\else
	  \'{c}\fi{}, Phys. Rev. B {\bf 85},  054406  (2012).
	
	\bibitem{Manchon2011}
	A. Manchon, Phys. Rev. B {\bf 83},  172403  (2011).
	
	\bibitem{Zhao2010}
	E. Zhao, C. Zhang, and M. Lababidi, Phys. Rev. B {\bf 82},  205331  (2010).
	
	\bibitem{Butch2010}
	N.~P. Butch {\it et~al.}, Phys. Rev. B {\bf 81},  241301  (2010).
	
	\bibitem{Qi2008}
	X.-L. Qi, T.~L. Hughes, and S.-C. Zhang, Nat Phys {\bf 4},  273  (2008).
	
	\bibitem{Mahfouzi2010}
	F. Mahfouzi, B.~K. Nikoli\'{c}, S.-H. Chen, and C.-R. Chang, Phys. Rev. B {\bf
	  82},  195440  (2010).
	
	\bibitem{Mahfouzi2016}
	F. Mahfouzi, B.~K. Nikoli\ifmmode~\acute{c}\else \'{c}\fi{}, and N. Kioussis,
	  Phys. Rev. B {\bf 93},  115419  (2016).
	
	\bibitem{Keldysh1965}
	L. Keldysh, Sov. Phys. JETP {\bf 20},  1018  (1965).
	
	\bibitem{Sinova2014}
	H.~K. et~al, Nature Nanotechnology {\bf 9},  211  (2014).
	
	\bibitem{Haug2008}
	H. Haug and A.-P. Jauho, {\em Quantum kinetics in transport and optics of
	  semiconductors} (Springer-Verlag, Berlin, ADDRESS, 2008).
	
	\bibitem{Bode2012}
	N. Bode {\it et~al.}, Phys. Rev. B {\bf 85},  115440  (2012).
	
	\bibitem{Arrachea2007}
	L. Arrachea, Phys. Rev. B {\bf 75},  035319  (2007).
	
	\bibitem{Jauho1994}
	A.-P. Jauho, N.~S. Wingreen, and Y. Meir, Phys. Rev. B {\bf 50},  5528  (1994).
	
	\bibitem{Gaury2014}
	B. Gaury {\it et~al.}, Physics Reports {\bf 534},  1  (2014).
	
	\end{thebibliography}

\end{document}